\documentstyle[12pt,epsfig]{article}

\newcommand{\beq}{\begin{equation}}
\newcommand{\eeq}{\end{equation}}
\newcommand{\bea}{\begin{eqnarray}}
\newcommand{\eea}{\end{eqnarray}}

\newcommand{\gsim}{\lower.7ex\hbox{$
\;\stackrel{\textstyle>}{\sim}\;$}}
\newcommand{\lsim}{\lower.7ex\hbox{$
\;\stackrel{\textstyle<}{\sim}\;$}}

\def\lsim{\mathrel{\rlap{\lower3pt\hbox{\hskip0pt$\sim$}}
    \raise1pt\hbox{$<$}}}         
\def\gsim{\mathrel{\rlap{\lower4pt\hbox{\hskip1pt$\sim$}}
    \raise1pt\hbox{$>$}}}         

\newcommand{\bibit}[1]{\bibitem{#1}}

\newcommand{\aver}[1]{\langle #1\rangle}

\newcommand{\La}{\overline{\Lambda}}
\newcommand{\Si}{\overline{\Sigma}}
\newcommand{\Lam}{\Lambda_{\rm QCD}}

\newcommand{\tto}{\!\to\!}

\newcommand{\GeV}{\,\mbox{GeV}}
\newcommand{\MeV}{\,\mbox{MeV}}
\newcommand{\matel}[3]{\langle #1|#2|#3\rangle}
\newcommand{\state}[1]{|#1\rangle}

\newcommand{\msp}[1]{\mbox{\hspace*{#1mm}~}}

\begin{document}
\thispagestyle{empty}
\vspace*{-10mm}

\begin{flushright}
Bicocca-FT-04-12\\
UND-HEP-04-BIG\hspace*{2pt}07\\
hep-ph/0409125
\end{flushright}
\vspace*{12mm}

\begin{center}
{\LARGE{\bf Heavy quark expansion in beauty:\vspace*{3mm}\\
recent successes and problems
}}
\vspace*{10mm}

{\small {\tt Invited talk at}}  { ~{\it Continuous Advances in QCD
\,2004}} \hspace*{75pt}~\\
{\small {\sf \hspace*{20pt} 13-16 May 2004,  Minneapolis, MN USA}} 
\vspace*{12mm}
\end{center}

\begin{center}
{\LARGE Nikolai~Uraltsev}\hspace*{1.5pt}\raisebox{5.5pt}{*}
\vspace*{6mm} \\
{\sl INFN, Sezione di Milano,  Milano, Italy}
\vspace*{30mm}

{\bf Abstract}\vspace*{-.9mm}\\
\end{center}

\noindent
The status of the QCD-based heavy quark expansion
is briefly reviewed. A good agreement between properly applied
theory and new precision data is observed. Critical remarks on certain
recent claims from HQET are presented. Recent applications to the
exclusive heavy flavor transitions are addressed. The
`$\frac{1}{2}\!>\!\frac{3}{2}$' problem for the transitions into the
charm $P$-wave states is discussed.

\setcounter{page}{0}

\vfill

~\hspace*{-12.5mm}\hrulefill \hspace*{-1.2mm} \\
\footnotesize{
\hspace*{-12pt}\raisebox{3pt}{$^*$}On leave of absence from
Department of Physics, University of Notre Dame,
Notre Dame, IN 46556, USA \\
\hspace*{-10pt} and from 
St.\,Petersburg Nuclear Physics
Institute, Gatchina, St.\,Petersburg  188300, Russia}
\normalsize

\newpage

Heavy quark physics, in
particular electroweak decays of beauty particles, is now a well
developed field of QCD.
The most nontrivial dynamic predictions are 
made for sufficiently inclusive heavy
flavor decays admitting the local operator product expansion
(OPE). These predictions are phenomenologically important
-- they allow to reliably extract the underlying
CKM mixing angles $|V_{cb}|$ and $|V_{ub}|$ with record accuracy from
the data, or the fundamental parameters like $m_b$ and $m_c$. 
At the same time heavy quark theory yields informative dynamic 
results for a number of exclusive transitions as well. Recent years
have finally
witnessed a more united approach to inclusive and exclusive decays
which previously have been largely isolated. 
In this talk I closely follow the nomenclature of the review~\cite{ioffe}
where the principal elements of the heavy quark
theory can be found.

For a number of years there has been a wide spread opinion that the
predictions of the dynamic QCD-based theory were not in agreement
with the data, a sentiment probably still felt today by many.
The situation, in fact, has changed over the past few years. 
A better, more robust 
approach to the analysis has been put forward~\cite{amst}, made 
more systematic~\cite{slcm} and
applied in practice~\cite{delphi,babarprl}. 
The perturbative corrections for all inclusive semileptonic
characteristics have finally been calculated~\cite{slsf,trott}.
Experiments have
accumulated data sets of qualitatively better statistics and
precision.

Critically reviewing the status of the theory when
confronted with the data, we find that the formerly 
alleged problems are replaced by impressive agreement. Theory
often seems to work even better than can realistically be expected,
when pushed to the hard extremes. Old problems are left in the
past.

\section{Inclusive semileptonic decays: theory vs.\ data}

The central theoretical result~\cite{buvbs} 
for the inclusive decay rates of heavy
quarks is that they are not affected by nonperturbative physics at the
level of $\Lam/m_Q$ (even though hadron masses, and, hence the phase
space itself, are), and the corrections are given by the local heavy
quark expectation values -- $\mu_\pi^2$ and $\mu_G^2$ to order
$1/m_Q^2$, etc. Today's theory has advanced far beyond that and 
allows, for instance, to aim at an 1\% accuracy in $|V_{cb}|$ 
extracted from $\Gamma_{\rm sl}(B)$. A similar approach to $|V_{ub}|$ is more
involved since theory has to conform with the necessity for experiment
to implement significant cuts which discriminate against 
the $b\tto c\,\ell\nu$
decays. Yet the corresponding studies are underway and a 5\% accuracy
seems realistic.

There are many aspects theory must address to target this level of
precision. One facet is perturbative corrections, a subject of
controversial statements for many years. The reason goes back to
rather subtle aspects of the OPE. It may be partially elucidated by
Figs.~1 which shows the relative weight of gluons with different momenta
$Q$ affecting the total decay rate and the average hadronic recoil
mass squared $\aver{M_X^2}$, respectively. The contributions in the
conventional `pole'-type perturbative approach have long tails
extending to very small gluon momenta below $500\MeV$, especially for
$\aver{M_X^2}$; the QCD coupling $\alpha_s(Q)$ grows uncontrollably
there. These tails would be disastrous for precision calculations manifest,
for instance, through a numerical havoc once higher-order corrections
are incorporated. Yet applying literally the Wilsonian prescription
for the OPE with an explicit separation of scales in all strong
interaction effects, including the perturbative contributions,
effectively cuts out the infrared pieces! Not only do the higher-order
terms emerge suppressed, even the leading-order corrections become
small and stable. This approach, applied to heavy quarks long 
ago~\cite{upsetblmopefive} implies that the precisely defined running
heavy quark parameters $m_{b}(\mu)$, $\La(\mu)$,
$\mu_\pi^2(\mu)$, ... appear in the expansion, rather than ill-defined
parameters like pole masses, $\La$, $-\lambda_1$ employed by
HQET. Then it makes full sense to extract these genuine QCD
objects with high precision.

\begin{figure}[hhh]
\vspace*{-15pt}
\begin{center}
\mbox{\psfig{file=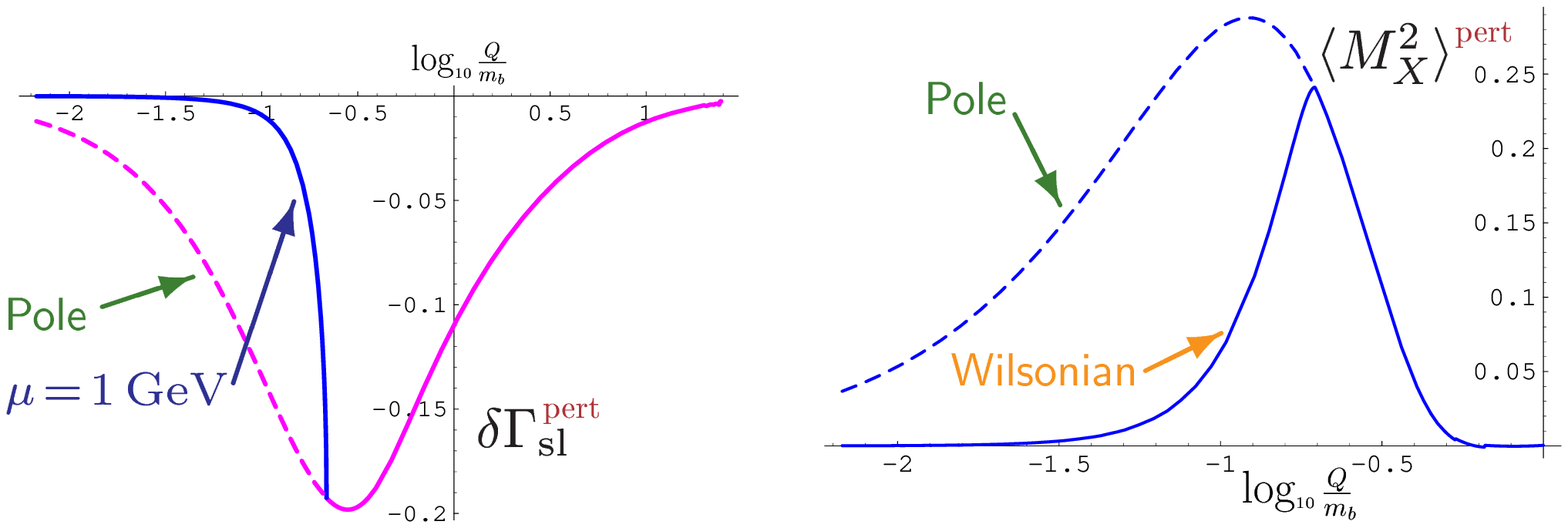,width=400pt}}\vspace*{-10pt}
\end{center}
\caption{{\small 
The role of the gluons with different momenta  
in $\Gamma_{\rm sl}$ and in $\aver{M_X^2}$, 
\,for $\,b\tto c\,\ell \nu$.}\vspace*{-5pt}} 
\end{figure}

\noindent
\begin{minipage}[t]{73mm}
\hspace*{1.2em}The most notable of all the alleged problems for the OPE in the
semileptonic decays was, apparently, the
dependence of the final state invariant hadron mass on the lower cut
$E_{\rm cut}^\ell$ in the lepton
energy: theory seemed to fall far off~\cite{ligman} of the
experimental data, see Fig.~2. The robust approach, on the contrary 
appears to describe it well~\cite{misuse}, as illustrated by
Figs.~3. The second moment of the same distribution also
seems to perfectly fit theoretical expectations~\cite{slcm,slsf}
obtained using 
the heavy quark parameters extracted by BaBar from their 
data~\cite{babarprl}. \linebreak
\end{minipage} 
\ ~\hfill \begin{minipage}[t]{75mm} \vspace*{-4mm}

~\hfill \mbox{\psfig{file=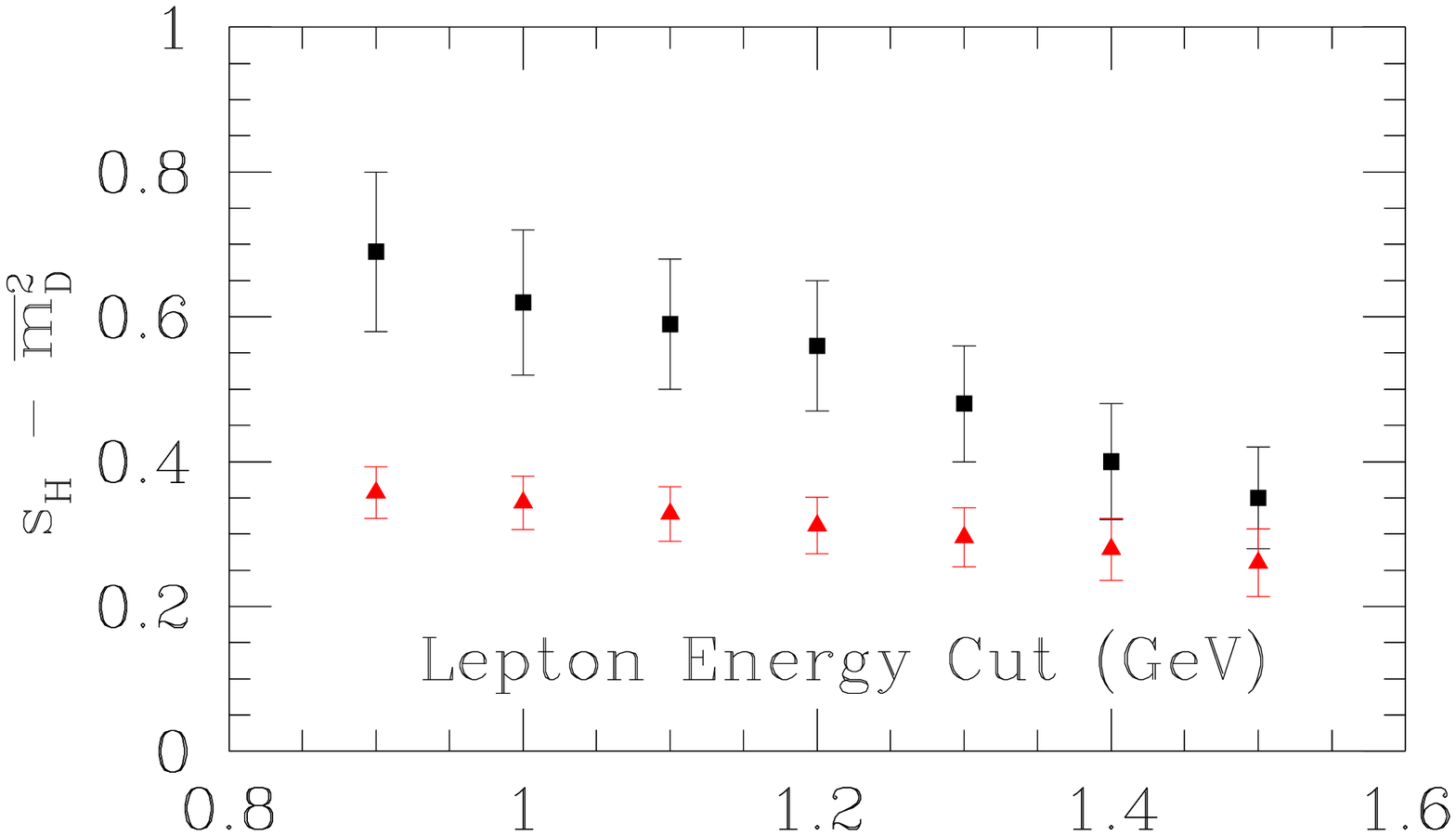,width=72mm}}\vspace*{-3.7mm}\\
{\small Figure\,\,2.~Ref.~\cite{ligman} 
predictions for $\aver{M_X^2}$  (red triangles), 
with the authors' theory error bars. Black squares are 
preliminary (2002) BaBar data points.\vspace*{-0mm} 

} 
\end{minipage}
\addtocounter{figure}{1}
\vspace*{-8.5pt}

\noindent
The second moment in the `inapt' calculations by Bauer 
{\it et al.}~\cite{ligman}, on the contrary showed unphysical growth with the
increase of $E_{\rm cut}^\ell$ in clear contradiction with
expectations and data.

\begin{figure}[hhh]
\vspace*{-9pt}
\begin{center}
\hspace*{-10pt}
\mbox{\psfig{file=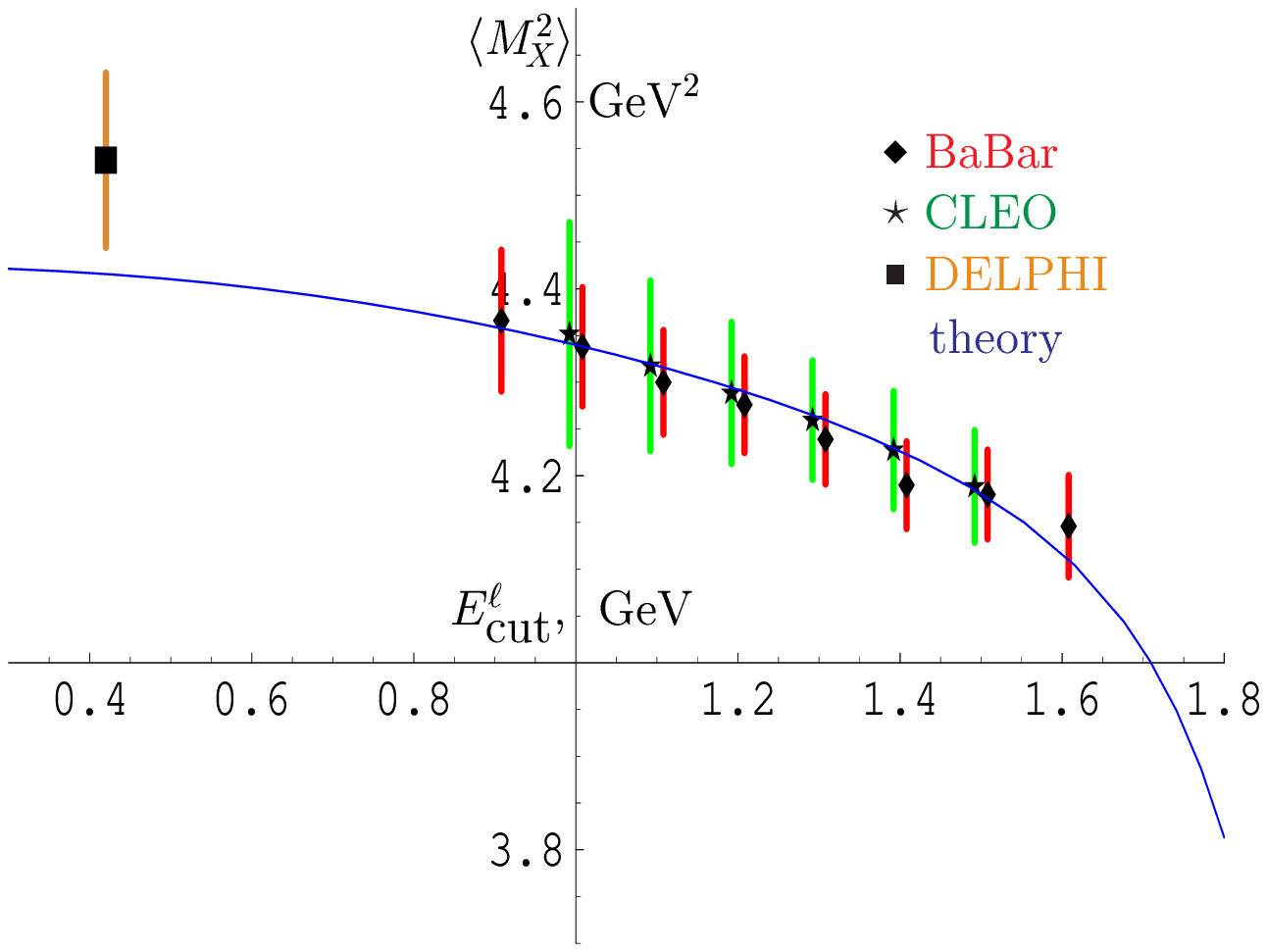,width=180pt}}\hspace*{40pt} 
\mbox{\psfig{file=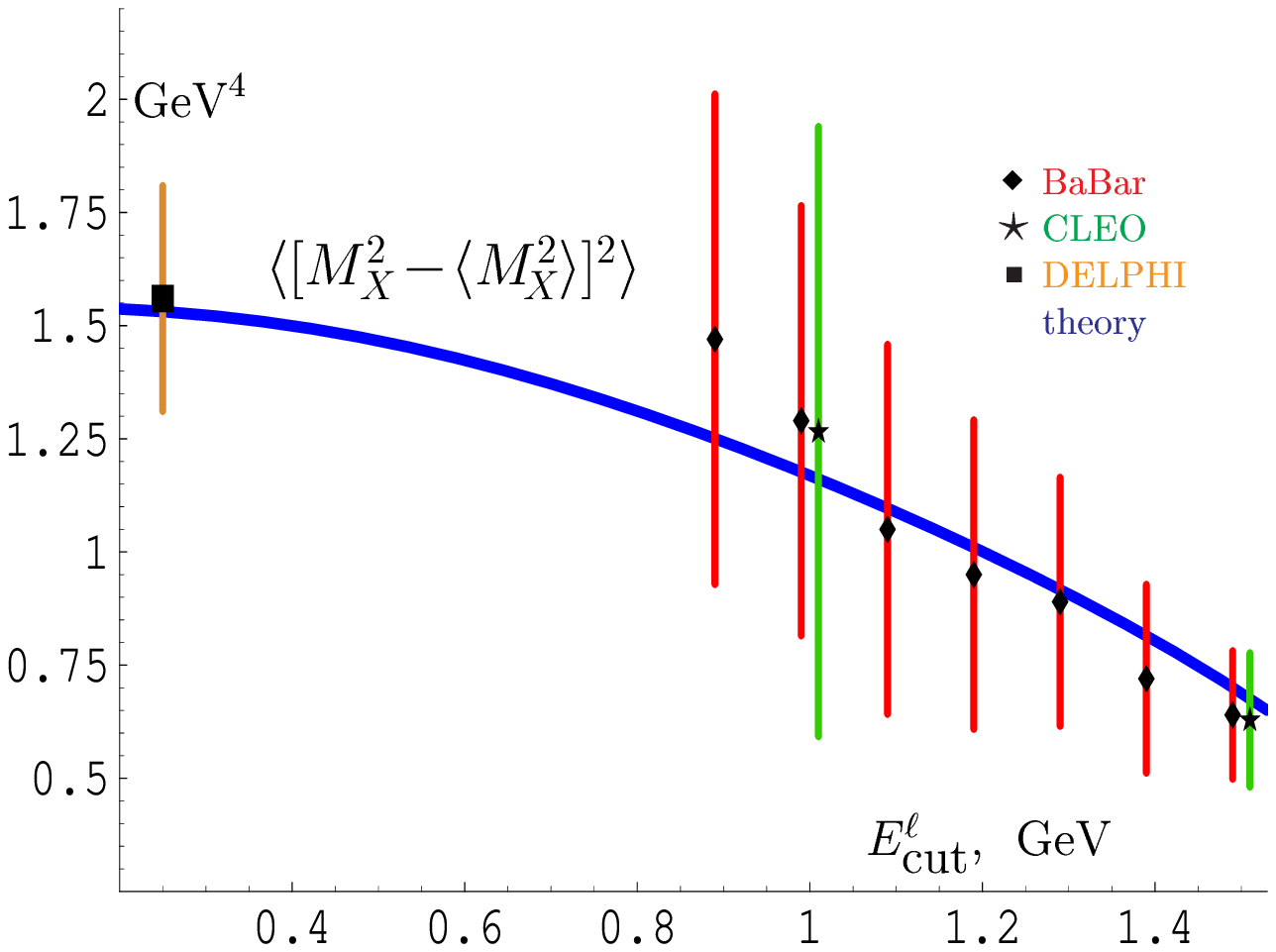,width=180pt}}\vspace*{-10pt}
\end{center}
\caption{{\small Hadron mass moments dependence on the lepton energy cut.} 
\vspace*{-12pt}}
\end{figure}

The comprehensive data analysis is now in the hands of professionals
(experimentalists) armed with the whole set of the elaborated 
theoretical expressions. They are able to perform extensive 
fits of all the available data
from different experiments, and arrive at rather accurate values of
the heavy quark parameters, still observing a good consistency of
data with theory. 
A number of such analyses are underway~\cite{fitreview}.

Another possible discrepancy between  data and theory used to be an 
inconsistency between the values of the heavy quark parameters
extracted from the semileptonic decays and from 
the photon energy moments~\cite{ligbsg} in
$B\tto X_s\!+\!\gamma$. It has been pointed out, however~\cite{misuse},
that with relatively high experimental cuts on $E_\gamma$ the actual
`hardness' ${\mathcal Q}$ significantly degrades compared to $m_b$, thus
introducing the new energy scale with ${\mathcal Q}\!\simeq\! 1.2\GeV$ at
$E_{\rm cut}^{\gamma} \!=\! 2\GeV$. Then the terms exponential in ${\mathcal
Q}$ left out by the conventional OPE, while immaterial under normal
circumstances, become too important. This is illustrated by Figs.~3
showing the related `biases' in the extracted values of $m_b$ and
$\mu_\pi^2$. Accounting for these effects appeared to turn
discrepancies into a surprisingly good agreement between all the
measurements~\cite{misuse}.

\begin{figure}[hhh]
\vspace*{-10pt}
\begin{center}
\mbox{\psfig{file=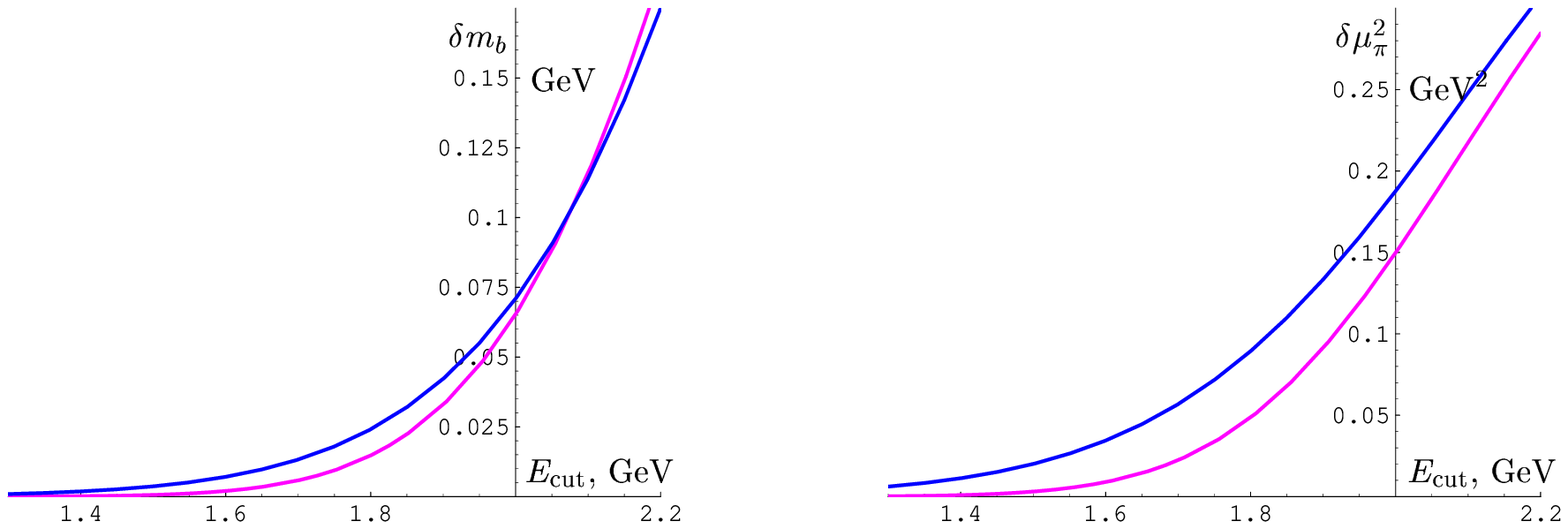,width=400pt}}\vspace*{-33pt} 
\end{center}
\caption{{\small 
`Exponential' in ${\mathcal Q}\,$ biases in $m_b$ and 
$\mu_\pi^2$ due to the lower cut on photon energy 
in $B\tto X_s+\gamma$.
}
\vspace*{-8pt}} 
\end{figure}

The problem of deteriorating hardness with high cuts and of the
related exponential biases raised in~\cite{misuse} was well 
taken by many experimental
groups. BELLE have done a very good job~\cite{belle} in pushing the cut on 
$E_\gamma$ down to $1.8\GeV$, 
which softens the uncertainties in the  biases:
\bea
\nonumber
\aver{E_\gamma}\;\; &\msp{-2}=\msp{-2}& 
2.292\;\:\pm 0.026_{\rm stat}\;\:\pm 0.034_{\rm sys}\; \GeV \\
\aver{(E_\gamma \!-\!\aver{E_\gamma})^2} &\msp{-2}=\msp{-2}& 0.0305
\pm 0.0074_{\rm stat}\pm 0.0063_{\rm sys} \GeV^2.
\label{40}
\eea
The theoretical expectations based on the central BaBar values of the
parameters with $m_b\!=\!4.612\GeV$,  $\mu_\pi^2\!=\!0.40\GeV^2$, for the
moments with $E_{\rm cut}^{\gamma}\!=\!1.8\GeV$
are 
\beq
\aver{E_\gamma} \simeq 2.316 \GeV\;,\qquad  
\aver{(E_\gamma -\aver{E_\gamma})^2}\simeq 0.0324 \GeV^2\;,
\label{42}
\eeq
again in a good agreement.

Although the heavy quark distribution
functions governing the shape of the decay distribution in the $b\tto
c$ and $b\tto u$ or $b\tto s$ transitions are different, the Wilsonian
OPE ensures that the nonperturbative part of the moments 
in all these decays is given by the {\sf same} heavy
quark expectation values. This fact appears very important in
practical studies aimed at extracting $|V_{ub}|$ from the inclusive
$B\tto X_u\,\ell\nu$ rates, since the accuracy in constraining the
heavy quark parameters achieved in the  $b\tto c\,\ell\nu$ measurements
is significantly higher than direct constraints from the radiative
decays. According to experimental analyses, incorporating the 
former information brings the currently achievable accuracy for
extracting $|V_{ub}|$ close to the $5\%$ goal.

As a brief summary, the data show good agreement with the properly
applied heavy quark theory. In particular, it appears that\\
$\bullet$ Many underlying heavy quark parameters have been accurately
determined directly from experiment.\\
$\bullet$ Extracting $|V_{cb}|$ from $\Gamma_{\rm sl}(B)$ has
high accuracy and rests on solid grounds.\\
$\bullet$ We have precision checks of the OPE-based theory at the
level where nonperturbative effects play the dominant role.

In my opinion, the most nontrivial and critical test for theory is the 
consistency found between the hadronic mass and the 
lepton energy moments, in particular $\aver{M_X^2}$ vs.\ $\aver{E_\ell}$. 
This is a sensitive check of the
nonperturbative sum rule for $M_B\!-\!m_b$, at the precision 
level of higher power corrections. It is interesting to note in this respect
that a particular combination of the quark masses, $m_b\!-\!0.74m_c$
has been determined in the BaBar analysis with only a $17\MeV$ error bar!
This illustrates how $|V_{cb}|$ can be obtained with high
precision: the semileptonic decay rate $\Gamma_{\rm sl}(B)$ is driven
by nearly the same combination~\cite{imprec}.

\subsection{Comments on the literature}

{\bf a) \,Semileptonic decays} ~The developed, 
thorough theoretical approach to the inclusive
distributions has not escaped harsh criticism from Ligeti {\it et al.}
which amounted to the strong recommendations not to use it for data
analysis, with the only legitimate approach assumed to be that of 
Ref.~\cite{ligman}. It was claimed, in particular that the observed correct
$E_{\rm cut}^\ell$-dependence of $\aver{M_X^2}$ is lost once the
complete cut-dependence of the perturbative corrections is included,
being offset by the growth in the latter. We showed these claims were
not true: the perturbative corrections remain small for the whole
interval of $E_{\rm cut}$ up to $1.4\GeV$, and actually are 
practically flat, Fig.~1 of Ref.~\cite{slsf}. 
Moreover, the figure shows that these perturbative
corrections with the full $E_{\rm cut}$ dependence in the traditional 
pole scheme {\sf decrease} for larger $E_{\rm cut}$, in agreement 
with intuition.

The problems in the calculations of Ref.~\cite{ligman} have not been
traced in detail; its general approach has a number of vulnerable elements,
and the calculations themselves were not really presented. There are
reasons to believe they actually contained plain algebraic mistakes. 

There is a deeper theoretical reason to doubt the validity of the
approach adopted in Ref.~\cite{ligman} based on the so-called 
``$1S$'' scheme which is pushed for the analysis of $B$
decays somewhat beyond reasonable limits. In respect to the OPE
implementation, it differs little, if any from the usual pole
scheme. Only at the final stage are the observables, like the total width 
which depends on the powers of $m_b$, re-expressed in terms of the
so-called `$1S$' $b$ mass. The latter is basically 
$\frac{M_{\Upsilon(1S)}}{2}$ in perturbation theory. 
Since no `$\Upsilon(1S)$ $c$-quark mass' exists,
for $b\!\to\! c$ decays the scheme intrinsically relies on the pole mass
relations, in particular to exclude $m_c$ from consideration. 
Use of the $1/m_c$ expansion certainly represents a weak point whenever
precision predictions are required.

In fact, there is a more serious concern about the legitimacy of the
perturbative calculations in the `$1S$' scheme, whose working tool
is the so-called `Upsilon expansion'~\cite{ups}. Surprisingly, it is
often not appreciated that this is not the conventional perturbative
expansion based on the algebraic rules for the usual power series in
an expansion parameter like $\alpha_s(m_b)$.
This framework rather involves more or less arbitrary
manipulations with the conventional perturbative series, referred to as a
`{\sf modified}' perturbative expansion. The rationale behind such
manipulations is transparent: the Coulomb binding energy of two
massive objects starts with $\alpha_s^2$ terms, hence $m_b^{1S}$
differs from the usual pole mass only to the  {\sf second order} in 
$\alpha_s$ :
\beq
m_b^{1S}= m_b^{\rm pole}\left[ 1- C_F^2\mbox{$\frac{\alpha_s^2}{8}$} + 
{\mathcal O}\left(\alpha_s^3, \beta_0\alpha_s^3 \ln{\alpha_s}\right)\right]\;.
\label{120}
\eeq
The last IR divergent terms $\propto \alpha_s^3 \ln{\alpha_s}$ simply
signify that the Coulomb bound state of a heavy quark $Q$ has an 
intrinsically different, lower momentum scale
$\alpha_s m_Q$; in a sense, this scale is zero in the conventional
perturbative expansion which assumes a series expansion 
around $\alpha_s\!\to\!
0$. Hence the relation is not infrared-finite, in the conventional
terminology. On the other hand, since the leading, 
${\mathcal O}(\alpha_s^2)$ term in Eq.~(\ref{120})
comes without $\beta_0$, the `$1S$' $\,b$ quark mass in all
available perturbative applications to $B$ decays has to be equated 
with the pole mass
$m_b^{\rm pole}$, whether or not a few  BLM corrections are
included. 

To get around this obvious fact, the `$\Upsilon$ expansion' postulated
considering a number of terms appearing to the $k$-th order in
perturbation theory, $c_k \alpha_s^k(m_b)$, to be actually of a lower
order, $c_n \alpha_s^n(m_b)$ with $n\!<\!k$. Since the power of the strong
coupling is explicit, this is done by introducing an ad hoc factor
$\epsilon\!=\!1$, making use of the property that unity remains unity
raised to arbitrary power. This ad hoc reshuffling constitutes the
heart of the `$\Upsilon$ expansion' and of using the `$1S$' $\,b$ quark mass
$m_b^{1S}$ in $B$ decays.\footnote{The original paper~\cite{ups}
presented some arguments calling upon the so-called ``large $n_f$
expansion'' supposed to justify reshuffling the orders. I
believe that the reasoning was wrong {\sf ab initio} missing the
basics of the renormalon calculus~\cite{revis,probe}.}

The scale $\,\alpha_s m_Q\,$ naturally appears in {\sf bound-state}
problems for heavy quarks since the perturbative
expansion parameter for nonrelativistic particles is not necessarily
$\alpha_s$, but rather runs in powers of $\alpha_s/v$, where $v$ is
their velocity. The analogue of the `bound-state' mass then 
naturally appears there, since powers of velocity make up for 
the missing powers of $\alpha_s$. 
Yet nothing of this sort is present in $B$ mesons
or in their decays, the $\epsilon$ parameters introduced by the 
`$\Upsilon$ expansion' is unity and can be placed ad hoc at any arbitrary
place. One clearly should not make up for the numerically larger than
$\alpha_s(m_b)$ value of
$\alpha_s(Q)$ at the smaller momentum scale $Q\!=\!\alpha_s m_b$ 
by equating at will 
terms of explicitly different orders in $\alpha_s$ in the usual
perturbative expansion.
The `$\Upsilon$ expansion' would be meaningless already in the
simplest toy analogue of the $B$ decays, muon $\beta$-decay. It is
then difficult to count on this approach to be sensible for
more involved $B$ decays where real OPE has to be used for high
precision.
\vspace*{1mm}

More recently, when this contribution was in writing, the new
paper by Bauer {\it et al.}\ appeared~\cite{ligmannew}. Claiming now
to describe the cut-dependence of $\aver{M_X^2}$, this paper came up with
new statements aimed at discrediting the Wilsonian approach and its
implementation. The authors assert that the approach we follow 
suffers from a large
`scale-dependence' when varying the Wilsonian separation scale
$\mu$. In addition, the authors state they cannot reproduce the
hadronic moments calculated in Refs.~\cite{slcm,slsf} used by
experimental groups for the data analysis. Once again I have to refute
the criticism -- the $\mu$-dependence turns out weak, actually far
below the expected level, as illustrated, for instance by Figs.~5 for
$\aver{E_\ell}$ and $\aver{M_X^2}$. The change in the moments from
varying $\mu$ 
corresponds to the variation in, say $m_b$ of only $4\MeV$ and
$1\MeV$, respectively! (Ref.~\cite{slcm} allowed an uncertainty of
$20\MeV$ due to uncalculated higher-order perturbative corrections.)

\begin{figure}[hhh]
\vspace*{-12pt}
\begin{center}
\hspace*{-10pt}
\mbox{\psfig{file=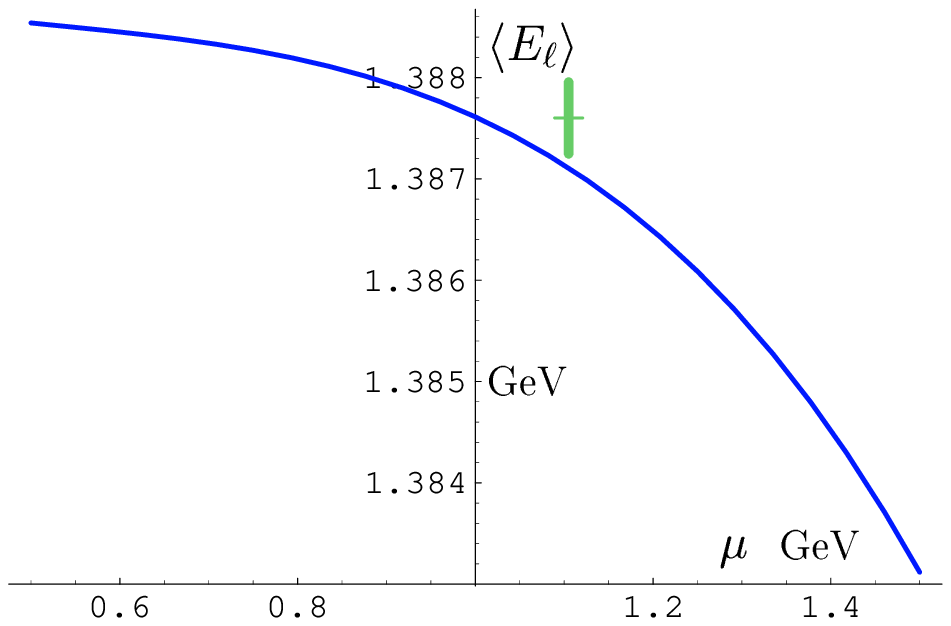,width=180pt}}\hspace*{60pt} 
\mbox{\psfig{file=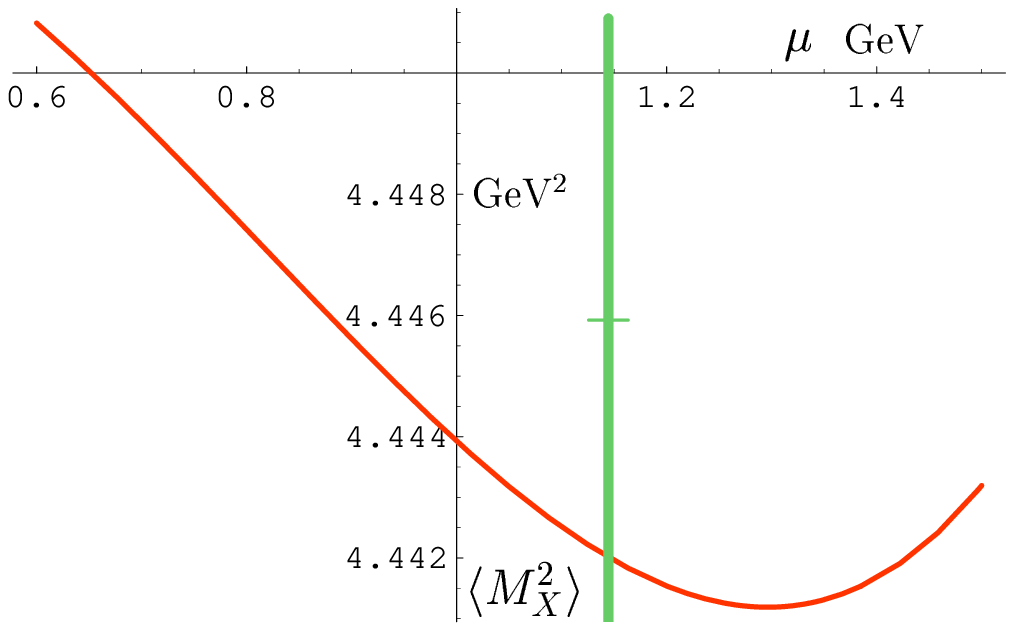,width=180pt}}\vspace*{-12pt}
\end{center}
\caption{{\small 
Dependence of $\,\aver{E_\ell}$ (left) and of 
$\,\aver{M_X^2}$ (right) 
on the separation scale $\mu$. The green vertical bars show the
change in the moments when $m_b$ is varied by $\pm 1\MeV$}
\vspace*{-1pt}}
\end{figure}

It looks probable that the authors of Ref.~\cite{ligmannew} simply were
not able to perform correctly the calculations in the Wilsonian
`kinetic' scheme, at least for the hadronic mass moments. In fact, the
suppressed dependence of the observables on the separation scale $\mu$ is
a routine check applied to the calculations. The two facts together are 
then rather suggestive.

Varying $\mu$ represents a useful -- if limited -- probe of the
potential impact of the omitted higher-order corrections.
Clearly, not varying $\mu$ but fixing
its value once and for all, one does not see any scale-dependence (the
pole scheme simply amounts to setting $\mu\!=\!0$). In this respect 
hints at an absent $\mu$-dependence in
the pole-type schemes like `$1S$' smell suspiciously. And, certainly, 
the absence 
of an explicit separation scale is not an advantage.

The analogous sensitivity to the actually used scale is of
course present in the approach of 
Refs.~\cite{ligman,ligmannew}, and the related uncertainties 
can be easily revealed. The `$1S$' scheme ad hoc postulates
using $m_b^{1S}$, a half of the $\Upsilon(1S)$ mass. However, on the
same grounds $m_b^{\eta_b}$,  half of the mass of the 
ground-state 
bottomonium, $\eta_b(1S)$ can be used. Even accepting the arbitrary
counting rules of the `$\Upsilon$-expansion', all the theoretical
expressions used in the analyses, are identical for  $m_b^{\eta_b}$ and
$m_b^{1S}$ -- the masses differ only to order $\alpha_s^3$ (without
$\beta_0$). At the same time, the two $b$ quark masses do 
differ numerically by at least $20$ to $30\MeV$! This is significantly
larger than the criticized $\mu$-dependence of the `kinetic' Wilsonian
scheme, and it, in any case, should be included as the minimal theory
uncertainty of every calculation based on the `$\Upsilon$'-mass of the
$b$ quark (it has not been, of course).

The theory error estimates of Refs.~\cite{ligman,ligmannew}, upon
inspection, look unrealistic, significantly underestimating
many potential corrections. The numerical outcome of the fit for $|V_{cb}|$
looks close to the value obtained by experimental groups in our
approach, within the error estimates we believe are right.  This
impression would be superficial -- the two calculations share many
common starting assumptions; therefore, they must yield -- if
performed correctly -- much closer results. One can state they do
differ at a level which is significant theoretically.

In this respect I would urge experiments to refrain from averaging the
results obtained in the two approaches. It is never a good idea to
combine correct results with those based on a potentially flawed
calculations. In my opinion, those relying on the `Upsilon expansion'
can be considered as such. For instance, the authors of
Ref.~\cite{ligman}, according to  their Eq.~(26) and Table~I increase the
value of $|V_{cb}|$ due to electroweak corrections (the same appears to
apply to the recent Ref.~\cite{ligmannew}). The fact is the
electroweak factor $\eta_{\mbox{{\tiny {\sf QED}}}}$ 
increases the width and, therefore
{\tt suppresses} the extracted value of $|V_{cb}|$ by an estimated
$0.7\%$. 

It is curious to note that, assuming this is just a mistake rather
than yet another ad hoc postulate of the `Upsilon expansion', correcting
for it would make the $|V_{cb}|$ value of Ref.~\cite{ligmannew}
nearly identical to the result obtained by BaBar. Whether such
a correspondence is inevitable, or is a matter of coincidence, is not
obvious at the moment. The existing PDG reviews on the subject have been
so far based exclusively on the questionable papers ignoring
more thorough existed analyses, and may therefore represent a not too
trustworthy source of information.
\vspace*{2mm}

\noindent
{\bf b) \boldmath $b\tto s\!+\!\gamma$ \,and\, $b\tto u\!+\!\ell\nu$} ~There
is a subtlety in accounting for the perturbative effects in the
heavy-to-light  decays which we do not see in $b\tto c\,\ell\nu$. The
radiated gluons can be emitted with sufficiently large energy yet at a
very small angle, so that their transverse momentum is only of order
$\mu_{\rm hadr}$ or even lower. This is a nonperturbative regime, and
it may generate a new sort of the nonperturbative corrections. These are
physically distinct from the Fermi motion encoded in the distribution
function of the heavy quark inside the $B$ meson. A dedicated
discussion can be found in the recent paper~\cite{jet}. Such
contributions may indicate that the so-called `soft-collinear
effective theories' (SCET), in all their variety, may
not truly represent an effective theory of actual QCD, not having
the identical {\sf nonperturbative} content.
It has been shown in Ref.~\cite{jet} that this physics, nevertheless
do not affect the moments of the decay distributions, in particular
the photon energy moments (at a low enough cut). The relation of the
moments to the local heavy quark expectation values remains unaltered:
the perturbative corrections have the usual structure and include only
truly short-distance physics. In this respect, we do not agree with
the recent claims found in the literature that the usual OPE relations for
the moments in the light-like distributions do not hold where
perturbative effects are included.

Our analysis does not support large uncertainties in the $b\tto
s+\gamma$ moments reported by Neubert at the Workshop, see also
Refs.~\cite{neubnew}. (It is curious to note the \,{\tt increase}\, in 
$\aver{E_\gamma}$ when lowering $E_{\rm cut}$ obtained by the
author). I would disagree already with the starting point of that approach. 
On the contrary, applying the Wilsonian approach we
find\,~\cite{bias} quite accurate, stable (and physical as well) 
predictions whenever the cut on
the photon energy is sufficiently low to cover the major part of the
distribution function domain.

\section{ A `BPS' expansion}

The heavy quark 
parameters as they emerge from the fit of the data are 
close to the theoretically expected values,  $m_b(1\GeV)\!\simeq\!
4.60\GeV$, $\mu_\pi^2(1\GeV)\!\simeq\!
0.45\GeV^2$, $\rho_D^3(1\GeV)\!\simeq\! 0.2\GeV^3$. The precise value, in
particular of $\mu_\pi^2$, is of considerable theoretical interest. It
is essentially limited from below by the known chromomagnetic
expectation value\,~\cite{ineq}:
\beq
\mu_\pi^2(\mu)> \mu_G^2(\mu), \qquad \mu_G^2(1\GeV)\simeq 
0.35^{+.03}_{-.02}\GeV^2,
\label{60}
\eeq
and experiment seem to suggest that this bound is not too far from
saturation. This is a peculiar regime where the heavy
quark sum rules~\cite{ioffe}, the exact relations 
for the transition amplitudes between sufficiently heavy flavor hadrons, 
become highly 
constraining. 

One consequence of the heavy quark sum rules is the lower 
bound~\cite{newsr} on
the slope of the IW function $\varrho^2\!>\!\frac{3}{4}$. They also 
provide upper bounds which turn out quite
restrictive once $\mu_\pi^2$ is close to $\mu_G^2$, \,say
\beq
\varrho^2 \!- \mbox{$\frac{3}{4}$} \lsim  0.3 \;\; \mbox{ ~if~ }\;\; 
\mu_\pi^2(1\GeV)\!-\! \mu_G^2(1\GeV) \lsim 0.1\GeV^2.
\label{64}
\eeq
This illustrates the power of the comprehensive 
heavy quark expansion in QCD: the
moments of the inclusive semileptonic decay distributions can tell 
us, for instance, about the formfactor for $B\tto D$ or $B\tto D^*$ decays.

Another application is the $B\tto D\,\ell \nu$ amplitude near zero
recoil. Expanding in $\mu_\pi^2\!-\!\mu_G^2$ an accurate estimate was
obtained~\cite{bps}
\beq
\frac{2\sqrt{M_B M_D}}{M_B+M_D} f_+(0)\simeq 1.04\pm 0.01\pm 0.01\;.
\label{70}
\eeq
In fact, $\mu_\pi^2\!\simeq\! \mu_G^2$ is a remarkable physical point for
$B$ and $D$ mesons, since the equality implies a functional
relation $\vec\sigma_b\vec\pi_b\state{B}\!=\!0$. Some of the Heavy
Flavor symmetry relations (but not those based on the spin
symmetry) are then preserved to {\sf all orders} in $1/m_Q$. This
realization led to a `BPS' expansion~\cite{chrom,bps} where
the usual heavy quark expansion was combined with an expansion around
the `BPS' limit $\vec\sigma_b\vec\pi_b\state{B}\!=\!0$. 

There are a number of miracles in the `BPS' regime. They  
include $\varrho^2\!=\!\frac{3}{4}$ and 
$\:\rho_{LS}^3\!=\!-\rho_D^3$; a complete discussion can be found in
Ref.~\cite{bps}. 
Some intriguing ones are~\cite{fpcp03}:\\
$\bullet$ No power corrections to the relation
$M_P\!=\!m_Q+\La$ and, therefore to $m_b\!-\!m_c=M_B\!-\!M_D$. \\
$\bullet$ For the $\;B\!\to\! D\;$ amplitude the heavy quark limit 
relation between
the two formfactors 
\beq
f_-(q^2)=-\frac{M_B\!-\!M_D}{M_B+M_D}\; f_+(q^2)
\label{106}
\eeq
does not receive power corrections.\\
$\bullet$ For the zero-recoil $\;B\!\to\! D\;$ amplitude all
$\,\delta_{1/m^k}\,$ terms vanish.\\
$\bullet$ For the zero-recoil formfactor $\,f_+\,$ controlling decays 
with massless leptons
\beq
f_+((M_B\!-\!M_D)^2)=\frac{M_B+M_D}{2\sqrt{M_B M_D}}
\label{108}
\eeq
holds to all orders in $1/m_Q$.\\
$\bullet$ At arbitrary velocity, power corrections in $\;B\!\to\! D\;$
vanish,
\beq
f_+(q^2)=\frac{M_B+M_D}{2\sqrt{M_B M_D}} \;\,\mbox{{\large$ 
\xi$}}\!\left(\mbox{$\frac{M_B^2+M_D^2-
\raisebox{.6pt}{\mbox{{\normalsize $q^2$}}}}{2M_BM_D}$}\right)
\label{110}
\eeq
so that the $\;B\!\to\! D\;$ decay rate directly yields 
the Isgur-Wise function $\xi(w)$.

Since the `BPS' limit cannot be exact in actual QCD, we need to
understand the accuracy of its predictions.  The dimensionless
parameter $\beta$ describing the deviation from the `BPS' limit is not tiny,
similar in size to the generic $\,1/m_{c}\,$ expansion parameter, and
relations violated to order $\beta$ may in practice be more of a
qualitative nature.  However, the expansion parameters like
$\mu_\pi^2\!-\!\mu_G^2 \propto \beta^2$ can be good enough. One can
actually count together powers of $1/m_c$ and $\beta$ to judge the
real quality of a particular heavy quark relation.  In fact, the
classification in powers of $\beta$ to {\tt all orders} in $1/m_Q$ is
possible~\cite{bps}.

Relations (\ref{106}) and (\ref{110}) for the $B\!\to\!D$
amplitudes at arbitrary velocity can get first order corrections in
$\beta$, and may be not very accurate.  Yet the slope $\varrho^2$ 
of the IW function differs from $\frac{3}{4}$ only at 
order $\beta^2$.
Some other important `BPS' relations hold up to order $\beta^2$:\\
$\bullet$ $M_B\!-\!M_D=m_b\!-\!m_c$ and $M_D=m_c\!+\!\La$ \\
$\bullet$ Zero recoil matrix element $\matel{D}{\bar{c}\gamma_0 b}{B}$
is unity up to ${\mathcal O}(\beta^2)$\\
$\bullet$ The experimentally measured $B\!\to\!D$ formfactor $f_+$ near
zero recoil receives only second-order corrections in $\beta$ to all
orders in $1/m_Q$:
\beq
f_+\left((M_B\!-\!M_D)^2\right) = \frac{M_B\!+\!M_D}{2\sqrt{M_BM_D}} \;\,
+ {\mathcal O}(\beta^2)\;.
\label{116}
\eeq
The latter is an analogue of the Ademollo-Gatto theorem for the `BPS'
expansion, and is least obvious. The `BPS'
expansion turns out more robust than the conventional $1/m_Q$ one
which does not protect the decay against the first-order corrections.

As a practical application, Ref.~\cite{bps} derived an accurate
estimate for the formfactor $f_+(0)$ in the $B\tto D$ transitions,
Eq.~(\ref{70}), incorporating terms through $1/m_{c,b}^2$. The largest
correction, $+3\%$ comes from the short-distance perturbative
renormalization; power corrections are estimated to be only about $1\%$.
\vspace*{2mm}

\section{The \,{\boldmath `$\frac{1}{2}>\frac{3}{2}$'}\, problem}

So far mostly the success story of the heavy quark
expansion for semileptonic $B$ decays has been discussed. 
I feel obliged to 
recall the so-called `$\frac{1}{2}\!>\!\frac{3}{2}$' puzzle
related to the question of saturation of the heavy quark sum
rules. It has not attracted due attention so far, although it had been
raised independently by two teams~\cite{rev,ioffe,orsaysig}
including the Orsay heavy quark group, and it has been around for quite
some time. A useful
review was recently presented by A.~Le Yaouanc~\cite{leya32}; here I
briefly give a complementary view. 

There are two basic classes of the sum rules in the Small Velocity, 
or Shifman-Voloshin (SV) heavy quark limit. First are the spin-singlet
sum rules which relate $\varrho^2$, $\La$, $\mu_\pi^2$, $\rho_D^3$,... to the
excitation energies $\epsilon$ and transition amplitudes squared 
$|\tau|^2$ for the $P$-wave states. 
Both $\frac{1}{2}$ and $\frac{3}{2}$ $P$-wave states, i.e.\ those where 
the spin $j$ of the light cloud is $\frac{1}{2}$ or $\frac{3}{2}$, 
contribute to these sum rules.

The second class are `spin' sum rules, they express similar relations
for $\varrho^2\!-\!\frac{3}{4}$, $\La\!-\!2\Si$,
$\mu_\pi^2\!-\!\mu_G^2$, etc.  These sum rules include only
$\frac{1}{2}$ states.

The spin sum rules strongly suggest that the
$\frac{3}{2}$ states dominate over $\frac{1}{2}$ states, having
larger transition amplitudes $\tau_{3/2}$. In fact, this automatically
happens in all quark models respecting Lorentz covariance and the
heavy quark limit of QCD; an example are the Bakamjian-Thomas--type 
quark models developed at Orsay~\cite{orsayqm}, or the covariant models on
the light front~\cite{cheng}.

The lowest $\frac{3}{2}$ $P$-wave excitations of $D$ mesons, $D_1$ and
$D_2^*$ are narrow and well identified in the data. They seem to
contribute to the sum rules too little, with
$|\tau_{3/2}|^2\!\approx\! 0.15$ according to Ref.~\cite{leib}. 
Wide  $\frac{1}{2}$ states denoted by 
$D_0^*$ and $D_1^*$ are possibly produced more copiously; they can, in
principle, saturate the singlet sum rules. However, the spin sum rules
require them to be subdominant to the $\frac{3}{2}$ states. The most
natural solution for all the SV sum rules would be if the lowest 
$\frac{3}{2}$ states with $\epsilon_{3/2}\simeq 450\MeV$ have
$|\tau_{3/2}|^2\approx 0.3$, while for the $\frac{1}{2}$ states 
$|\tau_{1/2}|^2\approx 0.07\:\mbox{to}\:0.12$ with
$\epsilon_{3/2}\approx  300\:\mbox{to}\:500\MeV$.

Strictly speaking, higher $P$-wave excitations can make up
for the wrong share between the contributions of the lowest states. This
possibility is disfavored, however. In most known cases 
the lowest states in a given channel tend to saturate
the sum rules with a reasonable accuracy.

It should be appreciated that the above sum rules are exact for heavy
quarks. Likewise, the discussed consequences rely on the
assumptions most robust among those we usually employ in dealing with
QCD. Therefore, the problem we examine is not how in practice
$\tau_{3/2}$ might turn out less than $\tau_{1/2}$. Rather it is 
why, in spite of the actual hierarchy 
between $\tau_{3/2}$ and $\tau_{1/2}$ the existing
extractions seem to indicate the opposite relation.

In fact, the recent pilot lattice study~\cite{becir32} indicated the
right scale for both $\tau_{3/2}$ and $\tau_{1/2}$ and, taken at face
value, showed a reasonable saturation of the spin sum rule by the
lowest $P$-wave clan. Similar predictions had been obtained in the
relativistic quark model from Orsay~\cite{orsayqm} and in the
light-cone quark models~\cite{cheng}.

Experimentally $\tau_{3/2}$ and $\tau_{1/2}$ can be extracted from
either nonleptonic decays $B \tto D^{**}+\pi$ assuming factorization
and the absence of the final state interactions, or directly from
their yield in $B\tto D^{**}\,\ell\nu$ decays. The former way suffers
from possible too significant corrections to factorization, in
particular for the case of excited charm states. Such decays also 
depend on the amplitude at the maximal recoil, kinematically most
distant from the small recoil we need the amplitude at; we know 
that the slopes of the formfactors are quite significant even 
with really heavy quarks.

The safer approach is the direct yield in the semileptonic decays.
The data interpretation is obscured, however by the significant
corrections to the heavy quark limit for charm mesons. For instance,
the classification itself over the light cloud angular momentum $j$
relies on the heavy quark limit. However, one probably needs a good
physical reason to have the hierarchy between the finite-$m_c$ heirs
of the $\frac{1}{2}$ and $\frac{3}{2}$ states inverted, rather than
only reasonably modified compared to the heavy quark limit. Yet, as
has been shown, these corrections are generally significant and may
noticeably affect the extracted  $|\tau_{3/2}|^2$.

It has also been routinely assumed that the slope of the formfactors
are similar for the $\frac{3}{2}$ and for the $\frac{1}{2}$ $D^{**}$,
something which is not expected to hold in QCD. The existing
models likewise predict a large slope for the $\frac{3}{2}$ mesons and a
moderate one for the $\frac{1}{2}$ states. This clearly enhances the
actual extracted value of $|\tau_{3/2}|^2$.

The experimental situation in respect to the wide $\frac{1}{2}$ charm 
states still remains uncertain. It cannot be excluded that their
actual yield is smaller, and at the same time it can be 
essentially enhanced compared to the large-$m_c$ limit.

To summarize, we do not have a definite answer to how this apparent
contradiction of theory with data is resolved. Considering all the
evidence, the scenario seems most probable where all the above
factors contribute coherently, suppressing the yield of the
$\frac{3}{2}$ states more than 
expected  and enhancing the production of the 
$\frac{1}{2}$ states. First of all, this refers to the size of the
power corrections in charm. Secondly, the effect of significant formfactor
slopes for the $\frac{3}{2}$ states. Finally, it seems
possible that the actual branching fraction of the $\frac{1}{2}$
$P$-wave states in the semileptonic decays would be eventually below
$1\%$ level. In my opinion it is important to clarify this problem. 
\vspace*{2mm}

\noindent
{\bf Conclusions.}~ The dynamic QCD-based theory of inclusive 
heavy flavor decays has finally undergone and passed critical
experimental checks in the semileptonic $B$ decays at the
nonperturbative level. Experiment finds consistent values of the heavy
quark parameters extracted from quite different measurements once
theory is applied properly. The heavy quark parameters emerge close to
the theoretically expected values.  The perturbative corrections to
the higher-dimension nonperturbative heavy quark operators in the OPE
have become the main limitation on theory accuracy; this is likely to
change in the foreseeable future.

Inclusive decays can also provide important information for a number
of individual heavy flavor transitions.  The
$B\tto D\,\ell\nu$ decays may actually be accurately treated. The
successes in the dynamic theory of $B$ decays put a new range of
problems in the focus; in particular, the issue of the saturation of the
SV sum rules requires close scrutiny from both theory and experiment.
\vspace*{.1mm}

\section*{Acknowledgments}

I am grateful to D.~Benson, I.~Bigi, P.~Gambino, M.~Shifman,
A.~Vainshtein, O.~Buchmueller and P.~Roudeau, for 
close collaboration and discussions.
This work was supported in part by the NSF under grant number
PHY-0087419.

\end{document}